# InAs-GaSb laser: Prospects for efficient THz emission

L.D. Shvartsman and B. Laikhtman

The Racah Institute of Physics, The Hebrew University of Jerusalem, Jerusalem,

91904, Israel.

**Abstract.** We suggest to use InAs/GaSb coupled quantum wells for THz lasing. In these heterostructures THz lasing is based not on intersubband but on interband transitions. Crucial advantages of this design in comparison with intersubband lasers are (i) a large value of the interband dipole matrix element and (ii) easier maintaining of population inversion. These advantages lead to a gain of two orders of magnitude higher than for intersubband lasing. Even higher gain can be obtained in special design InAs/GaSb W-structures where a hybridization gap of 1-3THz is formed and optical density of states is singular.

Any design of efficient semiconductor lasing structures addresses the following set of problems: an appropriate gap, large matrix element of photon emission, high reduced density of states and creation of population inversion. We are going to show that InAs-GaSb coupled quantum well (CQW) structures provide both the required gaps [1-6] for interband transitions and the most prospective combination the rest three key parameters mentioned above [7] for THz lasing. This statement became especially important in the light of recent developments in the field of MIR lasers based on GaSb that show the best up to now performance even in comparison with such a traditionally leading material as GaAs [8,9].

THz radiation has a very wide range of highly important applications in such fields of high importance as medical imaging, security, extra-sensitive biochemical spectroscopy, etc. Nevertheless, the actual progress in these fields is hindered by the absence of low cost reliable sources similar to visible and near infrared semiconductor lasers. Historically the best achievements in THz lasers are reported for quantum cascade lasers based on intersubband transitions in GaAs/AlGaAs quantum wells [10-14]. Nevertheless, as we show below, the theoretical prospect for gain InAs-GaSb CQW-based interband THz lasers is better by a couple of orders of magnitude.

The very possibility to get interband transitions in THz range is based on the fact that InAs and GaSb form a type II heterostructure with the overlap of the InAs conduction band and GaSb valence band in bulk reaching 150 meV[6]. Starting from this band structure one can get THz gaps by two scenarios:

(i) For narrow enough quantum wells (QWs) the overlap can be removed by size-quantization. The in-plane dispersion law in this case resembles letter V (electrons) or inverted letter V (holes) and further we call it V-structure (Fig.1a). The simplest V-structure that can be used for THz-lasing consists of adjacent InAs and GaSb wells imbedded in between AlSb that presents a good barrier for both electron and holes. Playing with the widths of InAs and GaSb wells it is possible to tune the separation between the ground electron level in InAs well (e1) and ground hole level in the GaSb well (hh1) from negative values of more than 100 meV to positive values of the same order of magnitude. That is, in principle, the separation between e1 and hh1 levels can be adjusted for generation of radiation from a few μm [15] to hundreds of μm (1 THz corresponds to 300 μm) and longer.[16]

(ii) For wider quantum wells the initial overlap is not removed, but the hybridization gap is formed. [1-4] This gap naturally lies in THz range. [1-4] The in-plane dispersion law in this case resembles letter W (electrons) or inverted W (holes) and further on we call it W-structure (Fig.1b).

The resulting advantage of CQW interband lasers is illustrated in Fig. 2 where the gain ratio interband InAs-GaSb CQW laser and regular GaAs-based intersubband laser for of 3 THz and T=4K is plotted as a function of the pumping current. In entire range of practical actuality theoretical promise of interband lasers is at least two orders of magnitude higher. The physical explanation of this phenomenon is rather transparent and based on direct gain calculation.

The calculated gain for the two cases is presented in the Eq.1-2. Eq.1 describes the gain based on intersubband transition in a three level scheme where electrons are pumped to level 3, the radiative transition takes place between levels 3 and 2, and level 2 is depopulated by electron relaxation to level 1. The population inversion ($n_3$-$n_2$) in rhs of Eq.1 depends on pumping $J$. For the second case of interband transition in the system of InAs/GaSb adjacent quantum well Eq.2 theoretical gain is positive for any non-zero pumping.

The results are respectively

$$g_{isb} = \frac{e^2}{m_0^2 \omega} \frac{\Gamma}{n_{eff} cL} \frac{4\pi^2 \tau}{\hbar} |P_{isb}|^2 (n_3 - n_2), \qquad (1)$$

$$g_{ib} = \frac{e^2}{m_0^2 \omega} \frac{\Gamma}{n_{eff} cL} \frac{2\pi m}{\hbar^2} |P_{ib}|^2 [f_e(\varepsilon_{e\omega}) + f_h(\varepsilon_{h\omega}) - 1] \vartheta(\hbar\omega - E_0) \qquad (2)$$

Here $m_0$ is the free electron mass, $\omega$ is the radiation frequency, $\Gamma$ is the optical confinement factor, $n_{eff}$ is the effective refraction index, $L$ is the width of the electron confinement region (in InAs/GaSb coupled well this is a sum of the widths of the electron and hole wells, $L=L_e+L_h$), and $m = m_e^{InAs} m_h^{GaSb} / (m_e^{InAs} + m_h^{GaSb})$ is the reduced electron – hole mass. Due to nearly parallel dispersion of the initial and final levels in the intersubband transition the gain is proportional to the population inversion at them and is limited by the scattering width of the levels $\hbar/\tau$, while the interband gain depends on the occupation numbers $f_{e,h}$. The electron $\varepsilon_{e\omega} = \hbar^2 k_\omega^2/2m_e^{InAs}$ and hole $\varepsilon_{h\omega} = \hbar^2 k_\omega^2/2m_h^{GaSb}$ energies are defined from the relation $E_0 + \varepsilon_{e\omega} + \varepsilon_{h\omega} = \hbar\omega$ where $E_0$ is the energy gap. The factor containing the occupation numbers depend on $J$. If $J$ is lower than the transparency threshold $J^*$ this factor is negative.

The first and the most appealing advantage of interband THz radiative transitions is a large matrix element. In intersubband transition the radiation matrix element, $P_{isb}$, is the matrix element of electron momentum between the initial and final envelop wave functions. A rough estimate of this matrix element is $\hbar/L$ where $L$ and the length scale characterizing the envelop wave functions. Typically this scale is about a few nm or larger. On the other hand, in interband transitions the radiative matrix element, $P_{ib}$, is the matrix element of electron momentum between Bloch functions of the conduction and valence bands. The length scale characterizing these functions is about the lattice constant $a \approx 0.5$ nm. A rough estimate of this matrix element $\hbar/a$ gives a value more than order of magnitude larger than that of the intersubband matrix element. In reality this ratio of matrix elements is somehow smaller, because of the essential existence of tunneling across InAs/GaSb interface.

An accurate calculation of the intersubband matrix element gives

$$P_{isb} = \int \varsigma_f(z)\left(-i\hbar\frac{d}{dz}\right)\varsigma_i(z)dz = -i\omega m_e z_{if}, \tag{3}$$

where $\zeta_i(z)$ and $\zeta_f(z)$ are electron envelop functions of the initial and final states, and $z_{if}$ is the dipole matrix element. The interband matrix element in InAs/GaSb coupled quantum wells contains two main contributions: electrons from InAs tunnel to GaSb and there recombine with holes and holes from GaSb tunnel to InAs and there recombine with electrons. As a result,

$$P_{ib} = \frac{\sqrt{2}\pi^2}{\sqrt{L_e L_h}} \left[ p_{cv}^{InAs} \frac{m_h^{InAs}}{m_h^{GaSb}} \frac{L_e}{(\kappa_h^2 L_e^2 + \pi^2)\kappa_h L_h} + p_{cv}^{GaSb} \frac{m_e^{GaSb}}{m_e^{InAs}} \frac{L_h}{(\kappa_e^2 L_h^2 + \pi^2)\kappa_e L_e} \right] \quad (4)$$

where $\kappa_e$ and $\kappa_h$ are the decay decrement of the electron wave function in the GaSb well and hole wave function in the InAs well. The intersubband momentum matrix elements are calculated according to the empirical relation $p_{cv} = \sqrt{E_p m_0 / 2}$ where $E_p$=21.11 eV in InAs and $E_p$=22.88 eV in GaSb [17]. The results of the calculation of the matrix elements for three different frequencies are given in Table 1. In the Eq.(4) for $P_{ib}$ we neglected the dependence on the in-plane wave vector that normally arises from an admixture of light holes [1,19]. For GaSb-InAs nanostructures designed for lasing at a few THz this dependence comes out to be so weak that it cannot change our qualitative conclusions.

According to Table 1 the ratio $|P_{ib}|^2/|P_{isb}|^2$ is 40.2, 19.8, and 12.7 for 1 THz, 2 THz and 3 THz respectively.

The ratio of the gains (1-2) can be estimated as

$$\frac{g_{ib}}{g_{isb}} \approx A \frac{|P_{ib}|^2}{|P_{isb}|^2} \frac{m/2\pi\hbar\tau}{n_3 - n_2}. \quad (5)$$

Here A~1 is a geometrical factor including mostly the ratio of widths of regions of effective carrier confinement and depending on particular design. Effective masses $m_e^{InAs}$ =0.026 and $m_h^{GaSb}$=0.25 give m=0.023. Assuming the relaxation time $\tau$=10$^{-12}$ s we obtain $m/2\pi\hbar\tau = 3\times 10^9$ cm$^{-2}$. The population inversion can be estimated from the balance equations that give

$$n_3 - n_2 = J\tau_3\left(1 - \frac{\tau_{21}}{\tau_{32}}\right), \qquad (6)$$

where $J$ is the pumping, $\tau_3 = \tau_{31}\tau_{32}/(\tau_{31}+\tau_{32})$ and $\tau_{kl}$ is the electron transition time from level $k$ to level $l$. According to this expression the shorter $\tau_{21}$ the larger the population inversion is. Typical relaxation time is around 10$^{-12}$ s and to make the population inversion equal $3\times 10^9$ cm$^{-2}$s$^{-1}$ it is necessary to have $J = 3\times 10^{21}$ cm$^{-2}$s$^{-1}$ that corresponds to electric current density 480 A/cm$^2$. That is at this rather significant current density the interband gain is larger than the intersubband gain by more than order of magnitude.

THz lasing in intersubband case is based mostly on the efficient selective depopulation of the lower subband in intersubband transitions. This is done with the help of optical phonons emission. But in THz range this mechanism has very limited selectivity. This results in second and the most crucial advantage of interband lasers where there is no need to employ optical phonons. Indeed, when the separation between the upper and lower subbands is of the order of 1 THz (4.14 meV) phonon emission from the upper subband is nearly as much effective as emission from the lower one. (If in Eq.(6) $\tau_{21} \approx \tau_{31}$ then $n_3 - n_2 \leq 0.18 J\tau_{32}$, $\tau_{21}$ is usually optical phonon emission time, while $\tau_{32}$ is related mostly to e-e scattering). Large values of gain ratio ~400 (5) are based on the fact that $\tau_3$ and $\tau_{21}$ in (6) are essentially close to each other. Various sophisticated designs of intersubband THz lasing structures attempt to break this closeness of $\tau_3$ and $\tau_{21}$. The systems with resonant tunneling supply an additions time-scale, the tunneling time. An analysis shows that when the intersubband gain is optimized this new time-scale is in between $\tau_{ee}$ and $\tau_{LO}$. As a result the intersubband gain (1) becomes higher by the factor of the order of 1. However, the interband gain remains nearly two orders of magnitude higher than the intersubband one (3-4) for V-structures with the parameters mentioned in Table 1.

As we have already mentioned above the transition between e1 and hh1 levels in Fig. 1a is not the only possibility that can be used for THz generation. The overlap of the conduction and valence band and tunneling across the interface leads to the formation of hybridization gap with W-like dispersion of carriers.[1] (Fig. 1b) The typical width of the hybridization gap is naturally around 3 THz and may be adjusted by variation of CQW parameters [18]. This gap can be used for lasing. The W-structure has an additional advantage compared to others. The minimum of the upper band dispersion and the maximum of the lower band dispersion are reached not at points but at entire lines in k-space. Because of this the optical density of states has a singularity,

$$v_{opt}(\omega) = \frac{2m\omega}{\pi\hbar\sqrt{(\hbar\omega)^2 - E_g^2}}, \qquad (6)$$

where $E_g$ is the width of the optical gap. Scattering smears the singularity, this smearing gives $v_{opt} = (m/\pi\hbar^2)\sqrt{\omega\tau}$ at the edge. For $\tau$=1 ps the corresponding estimation leads to further increase of the gain of W-structure based 1 THz lasing by $\sqrt{\omega\tau} \approx 2.5$ times compared to V-structures.

The efficiency of THz radiation can be increased even more by making use of its small energy compared to typical energy of barriers in quantum wells. This feature makes it possible to create cascade lasing in k-space. Compared to simple InAs/GaAs coupled quantum wells here InAs quantum well is replaced with AlInAs/InAs quantum well with a step. Electrons pumped on level e3 emit a photon going to level e2, then relax from there with emission of optical phonon to level e1 and then emit another photon going to level hh1 (Fig.1a). The separation between these levels can be controlled not only by the widths of the wells but also by the composition. That is, another possible structure is AlSb/(InAs)$_{1-z}$(AlSb)$_z$/(InAs)$_{1-x}$(AlSb)$_x$/GaSb/AlSb, where $z > x$. As an example, in Table 2 we present the parameters of the structures with composition AlSb/(InAs)$_{0.98}$(AlSb)$_{0.02}$/InAs/GaSb/AlSb for lasing frequency 1 THz, 2 THz, and 3 THz. The energy step in the electron well for this composition
is 41.4 meV.

In conclusion, we suggest a new type of THz lasers based on interband transitions in InAs-GaSb heterostructures. Potentially these lasers can increase the radiation power by more than order of magnitude compared to intersubband lasers based on GaAs-AlGaAs heterostructures currently used in THz region. The InAs-GaSb lasers seem specifically attractive because GaAs-AlGaAs THz lasers practically reached their maximal possible efficiency.

REFERENCES

1. B. Laikhtman, S. de-Leone, and L. D. Shvartsman, Solid State Commun. **104**, 257 (1997); S. de-Leon, L.D.Shvartsman, and B. Laikhtman, Phys.Rev.B **60**, 1861 (1999).

2. M. J. Yang, C. H. Yang, B. R. Bennett, and B. V. Shanabrook, Phys. Rev. Lett. **78**, 4613 (1997).

3. L. J. Cooper, N. K. Patel, V. Drouot, E. H. Linfield, and D. A. Ritchie, and M. Pepper, Phys. Rev. B **57**, 11915 (1998).

4. Yu. Vasilyev, S. Suchalkin, K. von Klitzing, B. Meltser, S. Ivanov, and P. Kop'ev, Phys. Rev. B **60**, 10636 (1999).

5. I. Lapushkin, A Zakharova, S. T. Yen and K. A. Chao, J. Phys.: Condens. Matter **16,** 4677 (2004); A. Zakharova, S. T. Yen, and K. A. Chao, Phys. Rev. B **64**, 235332 (2001).

6. D.M.Symons, M.Lakrimi, R.J.Warbuton, R.J.Nicholas, N.J.Mason, P.J.Walder, and M.I.Eremets, Semicond.Sci.Technol.**9**, 118 (1994); M. S. Daly, D. M. Symons, M. Lakrimi, R. J. Nicholas, N. J. Mason and P. J. Walker, Semicond. Sci. Technol. **11,** 823 (1996); M.Roberts, N.J. Mason, S.G. Lyapin, Y.C. Chung and P.C. Klipstein*,* Int. Conf. on Phys. Semicond.*,* Osaka (2000)

7. J.R.Meyer, I. Vurgaftman, R. Q. Yang, and L.R. Ram-Mohan, Electron. Lett., **32**, 45 (1996); I. Vurgaftman, J. R. Meyer, and L. R. Ram-Mohan, IEEE Photonics Technol. Lett., **9**, 170 (1997).

8. K. Mansour, Y. Qiu, C.J. Hill, A. Soibel and R.Q. Yang, Electron. Lett. 42, 1034 (2006).

9. L. Shterengas, G. L. Belenky, J. G. Kim and R. U. Martinelli, Semicond. Sci. Technol. **19,** 655 (2004); L. Shterengas and G. Belenky, M. V. Kisin, D. Donetsky, Appl. Phys. Lett. **90**, 011119 (2007).

10. A. Tredicucci, R. Köhler L. Mahler, H. E Beere, E. H. Linfield and D. A. Ritchie, Semicond. Sci. Technol. **20,** S333 (2005).

11. B. S. Williams, S. Kumar, Q. Hu, and J. L. Reno, Optics Express **13**, 3331 (2005); S. Kumar, B. S. Williams, Q. Hu, and J. L. Reno, Appl. Phys. Lett. **88**, 121123 (2006).

12. C. Worrall, J. Alton, M. Houghton, S. Barbieri. H. E. Beere, D. Ritchie, and C. Sirtori, Optics Express **14**, 171 (2006).

13. G. Scalari, C. Walther, J. Faist, H. Beere and D. Ritchie, Appl. Phys. Lett. **88**, 141102 (2006); C. Walther, G. Scalari, J. Faist, H. Beere and D. Ritchie, Appl. Phys. Lett. **89**, 231121 (2006); G. Scalari, L. Sirigu, R. Terazzi, C. Walther, M. I. Amanti, M. Giovannini, N. Hoyler, J Faist, M. L. Sadowski, H. Beere, D. Ritchie, L. A. Dunbar and R. Houdre, J. Appl. Phys. **101**, 081726 (2007).

14. M. A. Belkin, F. Capasso, F. Xie, A. Belyanin, M. Fischer, A. Wittmann, and J. Faist, Appl. Phys. Lett. **92**, 201101 (2008).

15. B. Laikhtman S. Luryi and G. Belenky, J. Appl. Phys. **90**, 5478 (2001).

16. J. D. Bruno, R. Q. Yang, J. L. Bradshaw, and J. T. Pham, D. E. Wortman, *Proc.SPIE* Vol. 4287 (2001).

17. G. Bastard, *Wave mechanics applied to semiconductor heterostructures* (Halsted Press, New York,1988).

18. Smadar de-Leon, B.Laikhtman, L.D.Shvartsman, J.Phys.: Condensed Matter **10**, 8715-8729 (1998).

19. G. Shechter, L. D. Shvartsman, and J. E. Golub, Superlat. Microstructures, **19**, 383 (1996); G. Shechter and L. D. Shvartsman, Phys. Rev. B **58**, 3941 (1998); R. Magri, L. W. Wang, A. Zunger, I. Vurgaftman and J. R. Meyer, Phys. Rev. B **61**, 10235 (2000).

TABLES

Table 1. Possible parameters of structures for 1, 2, and 3 THz generation.

| $\omega$ THz | $L_e$ nm | $L_h$ nm | $\kappa_e$ nm$^{-1}$ | $\kappa_h$ nm$^{-1}$ | $P_{ib}$ $10^{-21}$ g cm/s | $|P_{isb}|$ $10^{-21}$ g cm/s |
|---|---|---|---|---|---|---|
| 1 | 26.0 | 3.26 | 0.93 | 1.93 | 1.77 | 0.279 |
| 2 | 27.1 | 3.20 | 0.93 | 1.92 | 1.74 | 0.391 |
| 3 | 27.4 | 3.15 | 0.93 | 1.91 | 1.78 | 0.500 |

**Table 2.** Parameters of the structure with the composition AlSb/(InAs)$_{0.98}$(AlSb)$_{0.02}$/InAs/GaSb/AlSb. Here $\omega$ is the radiation frequency, $a$ is the width of the InAs well, $b$ is the width of the (InAs)$_{0.98}$(AlSb)$_{0.02}$ step, $L_h$ is the width of the GaSb well, $E_{e1}$ and $E_{hh1}$ are the quantization energies of e1 and hh1 levels, $\Delta_{12}$ and $\Delta_{23}$ are the energy separations between e1-e1 and e2-e3 level.

| $\omega$ | a | b | $L_h$ | $E_{e1}$ | $\Delta_{12}$ | $\Delta_{23}$ | $E_{hh1}$ |
|---|---|---|---|---|---|---|---|
| THz | nm | nm | nm | meV | meV | meV | meV |
| 1 | 26.0 | 66.6 | 1.94 | 12.59 | 29.88 | 4.10 | 141.5 |
| 2 | 27.1 | 44.5 | 1.88 | 11.82 | 29.97 | 8.31 | 146.5 |
| 3 | 27.4 | 34.9 | 1.84 | 11.63 | 30.04 | 12.40 | 150.8 |

FIGURE CAPTIONS

Fig. 1(a)

A V-dispersion where interband radiative transitions between levels e1 and hh1 are realized. In a step InAlAs-InAs well it is possible to add also intersubband radiative transition between levels e3 and e2 realizing the cascade in k-space. Electrons relax from level e2 to level e1 emitting optical phonon.

Fig. 1(b).

Carrier dispersion near hybridization gap in InAs/GaSb coupled quantum wells.

Fig. 2

The ratio of the gains for interband and intersubband transitions vs pumping current. $J^*$ is the transparency current for interband lasing. The calculation is made for V-structure, 3 THz at 4 K.

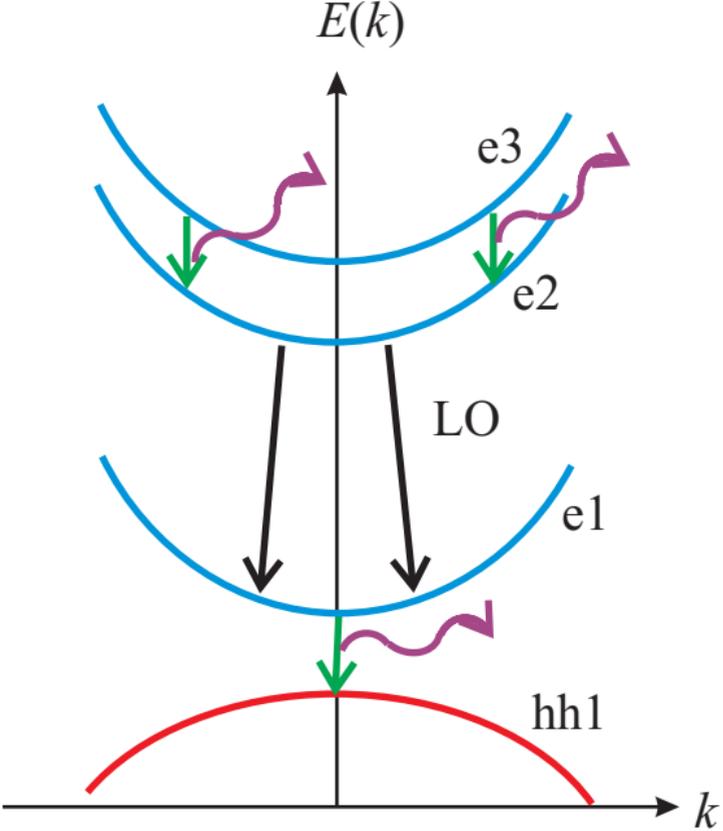 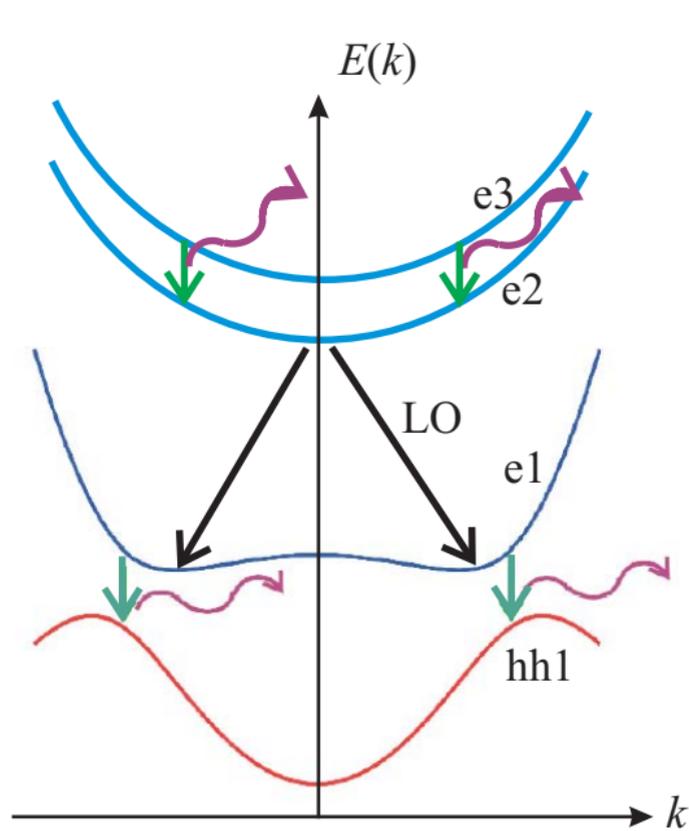

Fig. 1 (a)　　　　　　　　Fig. 1 (b)

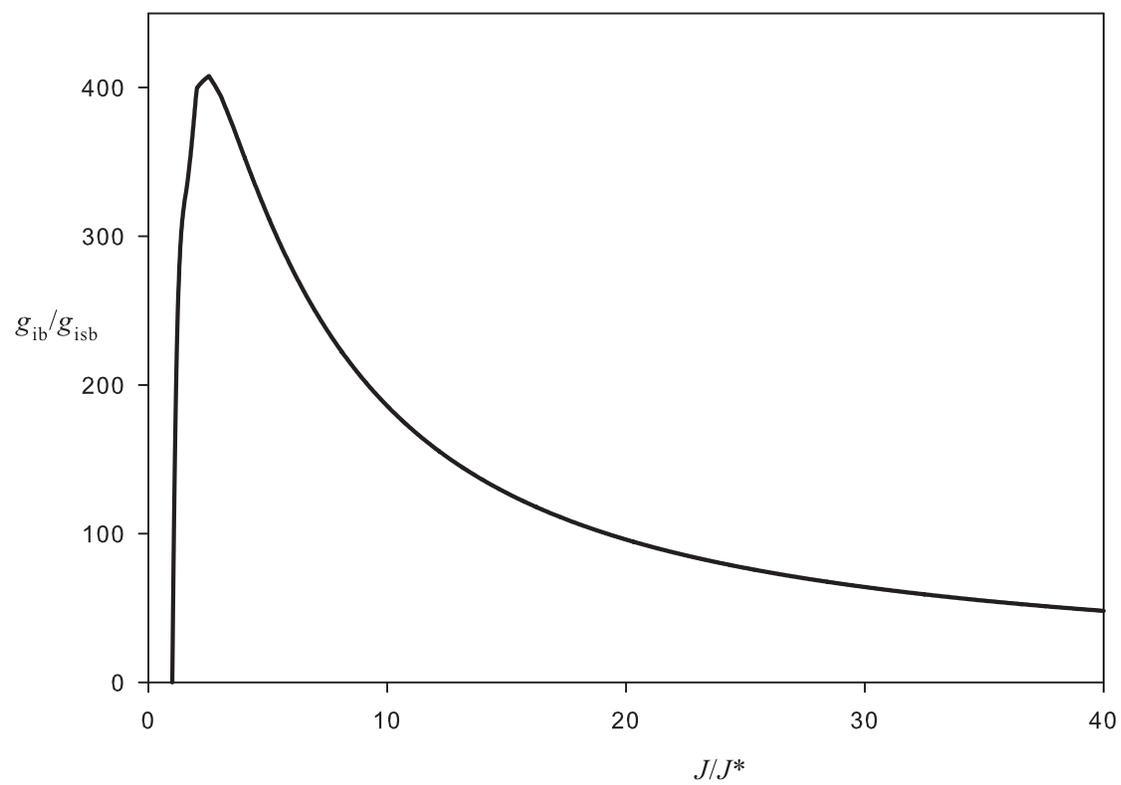